\documentclass[journal]{IEEEtran}
\usepackage[american]{babel}
\usepackage{amsmath,amssymb,amsfonts,mathrsfs}
\usepackage[amsmath,thmmarks]{ntheorem}
\usepackage{graphicx}
\usepackage{tikz}
\usetikzlibrary {positioning}
\usepackage{subfigure} 

\newtheorem{definition}{Definition}
\newtheorem{protocol}{Protocol}
\newtheorem{theorem}{Theorem}
\newtheorem{lemma}{Lemma}
\newtheorem{corollary}{Corollary}

\begin{document}
%
\title{Multi-User Non-Locality Amplification}
%
%
%

\author{Helen Ebbe and
        Stefan Wolf
\thanks{Manuscript submitted July 30, 2013. This work was supported by the Swiss National Science Foundation (SNF),
the NCCR "Quantum Science and Technology" (QSIT), and the COST
action on "Fundamental Problems in Quantum Physics." The results were presented in part at ISIT 2013 \cite{ebbwol}. This article has been submitted to IEEE Transactions on Information Theory and the copyright for the article has been transferred to IEEE.}
\thanks{H. Ebbe and S. Wolf are with the Faculty of Informatics, University of Lugano, 6900 Lugano,
Switzerland (e-mail: ebbeh@usi.ch; wolfs@usi.ch).}
}

\maketitle

\begin{abstract}
\emph{Non-local correlations} are among the most fascinating features
of quantum theory from the point of view of information: Such
correlations,
although not allowing for signaling, are unexplainable by pre-shared
information. The correlations have applications in cryptography,
communication complexity, and sit at the very heart of many attempts
of understanding quantum theory | and its limits | in terms
of classical information. In these contexts, the question is crucial
whether
such correlations can be \emph{amplified} or {\em distilled}, \emph{i.e.}, whether and how weak
correlations can be used for generating (a smaller amount of)
stronger.
Whereas the question has been studied quite extensively for \emph{bipartite}
correlations (yielding both pessimistic and optimistic results), only
little is known in the {\em multi-partite\/} case.

We introduce a general framework of reductions between multi-party input-output systems. Within this formalism, we show that a natural $n$-party generalization of the well-known
\emph{Popescu-Rohrlich box} can be distilled, by an adaptive protocol,
to the algebraic maximum. We use this result further to show that
a much broader class of correlations, including \emph{all} purely three-partite correlations, can be distilled from
arbitrarily weak to almost maximal strength with \emph{partial communication},
\emph{i.e.},
using only a subset of the channels required for the creation of the
same correlation from scratch. Alternatively, this means that arbitrarily
weak non-local correlations can have a ``communication value''
in the context of the generation of maximal non-locality.
\end{abstract}

\begin{IEEEkeywords}
Correlation distillation, information-theoretic systems, multiparty non-locality, quantum entanglement, quantum theory
\end{IEEEkeywords}

%
\IEEEpeerreviewmaketitle

\section{Introduction}
\IEEEPARstart{O}{ne} of the most mysterious, challenging, but also useful consequences
of quantum theory are non-local correlations: The
joint
behavior under (different possible) measurements of a quantum system
can be unexplainable by pre-shared (classical)
information
determining all the outcomes \emph{locally}. This result by Bell~\cite{bell}
can be seen as a late reply to the claim, in 1935, of Einstein,
Podolsky, and Rosen~\cite{EPR} that quantum theory was incomplete and
must be augmented by {\em hidden variables}, \emph{i.e.}, classical
information predicting all measurements' outcomes.\footnote{Bell's
 paradox only persists under the assumption that measurement bases are
chosen freely; at the same time, however, none of the \emph{deterministic}
interpretations of quantum physics satisfies with an \emph{explanation} neither of
the
correlations' origin nor of their limitations.}

It has been a prominent open problem why nature does
display non-local behavior, yet no maximal one. More specifically, why can Bell's inequality be violated, but a perfect \emph{Popescu-Rohrlich box}~\cite{pr} cannot be realized~\cite{cir}?
A number of attempts have been made to single out quantum correlations
among general non-signaling systems: Are quantum correlations
the ones that do not collapse {\em communication complexity\/}~\cite{cc}, that
are of no help for {\em non-local computation\/}~\cite{nlc}, that respect
{\em information causality}, a principle generalizing the non-signaling
principle to the case of limited communication~\cite{ic}, or that are \emph{locally orthogonal} \cite{locorth}, \emph{i.e.}, respect Specker's principle that if \emph{any pair} of questions about a system can be answered, then \emph{all questions together} can be answered simultaneously \cite{Cab12}?

It has turned out that non-local correlations have
important applications for information processing, \emph{e.g.},
device-independent cryptography or communication complexity. 
In all
these contexts, a question of paramount importance is the one
of
{\em distillation of non-locality\/}: Given weak correlations, is it
possible to generate stronger ones by local wirings?
For instance, distillation can potentially lead to higher
confidentiality levels or to a collapse of communication
complexity by (apparently) weak correlations.

In the two-party scenario, the possibility of distillation has already
been extensively studied and, notably, led to complementary results
adding up to a pretty complete picture: Whereas {\em isotropic
 CHSH-type}~\cite{CHSH}
correlations seem undistillable~\cite{dukwol}, the same fails to hold in general~\cite{FWW09}, \cite{BS09}, \cite{HoyerRashid}. In fact,
certain
arbitrarily weak CHSH correlations can even be distilled up to
virtually perfect PR boxes by adaptive protocols.

In the case of three or more parties, much less is known. It was shown
that the straight-forward generalization of the (non-adaptive) XOR
protocol~\cite{FWW09}
to more parties fails to distill extremal boxes of the non-signalling polytope to
almost-perfect~\cite{hsuwu}.

The contribution of the present work is as follows: We introduce a general framework for reductions of systems. In this model, we show that
the natural generalization of PR boxes to $n$ parties has the property
that non-isotropic faulty versions thereof can be distilled to
close-to-perfect
by a multi-party variant of Brunner and Skrzypczyk's \cite{BS09} protocol (Section IV). This result is
used to show distillability for a much larger class of correlations,
where the distillation is supported by partial communication, \emph{i.e.}, a
subset of the parties is allowed to communicate, whereas this communication
{\em alone\/} is insufficient for generating the target correlation
(Section~V). We call this partial communication supported distillation \emph{non-locality amplification}.
The result can alternatively be interpreted as arbitrarily weak
non-local correlations having a ``communication value'' in the context
of the generation of almost-perfect systems.
In Section~VI, the general results and procedures are illustrated with two examples.

\section{Systems, Boxes, and Non-Locality}
\subsection{Systems}
\begin{definition}[$n$-Partite System]
\emph{An \emph{$n$-partite system} is a conditional distribution
\begin{equation}
P_{A_1 A_2 \cdots A_n \vert X_1 X_2 \cdots X_n}\ ,
\end{equation}
where $X_i$ is the input and $A_i$ is the output variable of the $i$th party.}
\end{definition}
\subsection{Boxes are Non-Signalling Systems}
\begin{definition}[Non-Signaling]
\emph{\label{nsignaling}
An $n$-partite system with conditional probability distribution $P\left( a_1 a_2 \cdots a_n \vert x_1 x_2 \cdots x_n\right)$ is said \emph{non-signaling} if the marginal distribution for each subset of parties $\lbrace a_{k_1} , a_{k_2},..., a_{k_m}\rbrace$ only depends on its corresponding inputs
\begin{equation}
P\left( a_{k_1} \cdots a_{k_m}\vert x_1 \cdots x_n\right)  = P\left( a_{k_1} \cdots a_{k_m}\vert x_{k_1}  \cdots x_{k_m}\right)\ .
\end{equation}}
\end{definition}
An equivalent condition to Definition \ref{nsignaling} can be found in \cite{MAG06,BLMPPR05}:
\begin{eqnarray}
&\sum\limits_{a_k} P\left( a_1 \cdots a_k \cdots a_n \vert x_1 \cdots x_k \cdots x_n\right) =& \nonumber \\ &\sum\limits_{a_k} P\left( a_1 \cdots a_k \cdots a_n \vert x_1 \cdots x_k' \cdots x_n\right)&
\end{eqnarray}
for all $k\in \lbrace 1, 2,..., n\rbrace$, all inputs $a_1, a_2, ..., a_n$, and outputs $x_1, x_2, ..., x_{k-1}, x_k, x_k', x_{k+1}, ..., x_n$.
\begin{definition}[$n$-Partite Box]
\emph{An \emph{$n$-partite box} is a $n$-partite system that is non-signaling.}
\end{definition}
The ranges of $A_i$ and $X_i$, respectively, are arbitrary sets $\mathcal{A}_i$ and $\mathcal{X}_i$.

\subsection{Multipartite Locality}
Of central interest for us are $n$-partite boxes with the property that the parties cannot simulate the behavior of the box without communication but shared randomness only. This property is called \emph{non-locality}.
\begin{definition}[Local Box]
\emph{An $n$-partite box with input variables $X_1$, $X_2$, ..., $X_n$ and output variables $A_1$, $A_2$, ..., $A_n$ is local if
\begin{equation}\label{lokal}
P_{A_1 A_2 \cdots A_n|X_1 X_2 \cdots X_n}=\sum_{r\in{\cal R}}{P_R(r)\cdot P_{A_1|X_1}^r\cdots P_{A_n|X_n}^r}\
\end{equation}
for some random variable $R$.}
\end{definition}
Equivalently, there exists a distribution $P$ under which all joint outputs coexist. 

\begin{lemma}[Locality means Realism]
A box $P$ is local if and only if there exists a distribution
\begin{equation}
P'_{A_{1,0}A_{1,1}\cdots A_{1,\vert\mathcal{X}_1\vert -1}A_{2,0} \cdots A_{2,\vert\mathcal{X}_2\vert-1}\cdots A_{n,0}\cdots A_{n,\vert\mathcal{X}_n\vert -1}}
\end{equation}
with the property that its marginals satisfy
\begin{equation}\label{real}
P'_{A_{1,i_1}\cdots A_{n,i_n}}=P_{A_1 \cdots A_n|X_1=i_1, ..., X_n=i_n}
\end{equation}
for any $i_j\in\mathcal{X}_j$ for $j\in \{1, 2,..., n\}$. 
\end{lemma}

\begin{IEEEproof}
We assume that $P'$ exists and define the random variable 
\begin{equation}
R :=A_{1,0}A_{1,1}\cdots A_{1,\vert\mathcal{X}_1\vert -1}A_{2,0} \cdots A_{n,0}\cdots A_{n,\vert\mathcal{X}_n\vert -1}\ .
\end{equation} 
Obviously, this random variable $R$ satisfies (\ref{lokal}).

Assume that $P$ is local. In order to see that $P'$ exists, we define
\begin{eqnarray}
P'_{A_{1,0}A_{1,1}\cdots A_{n,0}A_{n,1}}(a_{1,0}a_{1,1}\cdots a_{n,0}a_{n,1}):= \nonumber\\
\sum_{r\in{\cal R}}{P_R(r)\cdot \prod_{i=1}^{n} P_{A_i|X_i}^r(a_{i,0},0)\cdot P_{A_i|X_i}^r(a_{i,1},1)}
\end{eqnarray}
and compute the marginals.

\end{IEEEproof}

Throughout, the remainder of this article, all the ranges   $\mathcal{A}_i$ and $\mathcal{X}_i$ are assumed to be $\{0, 1\}$.

\subsection{Specific Non-Local Boxes}
We define certain classes and specific types of $n$-partite boxes which we will use for our reductions. They are generalizations of the bipartite boxes studied in \cite{FWW09,BS09,BP05}.

We focus our attention to \emph{full-correlation boxes}. Intuitively speaking, such a box displays correlation only with respect to the \emph{full} set of players.

In the following definitions, the $n$-tuple of inputs is denoted by $\textit{\textbf{x}} = (x_1,x_2,...,x_n)$, where $x_i \in \lbrace 0, 1\rbrace$. The $n$-tuple of outputs is $\textit{\textbf{a}} = (a_1,a_2,...,a_n)$, where $a_i \in \lbrace 0, 1\rbrace$ for all $i$. 
\begin{definition}[Full-Correlation Box]
\emph{An $n$-partite \emph{full-corre\-lation box} is characterized by the following conditional distribution:
\begin{equation}
P(\textit{\textbf{a}}\vert \textit{\textbf{x}}) = \begin{cases} \frac{1}{2^{n-1}}&\text{$\sum\limits_i a_i \equiv f(\textit{\textbf{x}})$ (mod 2)}\\0&\text{otherwise,}\end{cases}
\end{equation}
where $f(\textit{\textbf{x}})$ is a Boolean function of the inputs. }
\end{definition}

Two special cases of the full-correlation boxes are the \emph{$n$-partite Popescu-Rohrlich box} and the \emph{even-parity box for $n$ parties}.
\begin{definition}[$n$-Partite Popescu-Rohrlich Box]
\label{def:pr}
\emph{An \emph{$n$-par\-tite Popescu-Rohrlich box} (or \emph{n-PR box}) is characterized by the following conditional distribution
\begin{equation}
P^{\text{PR}}_n(\textit{\textbf{a}}\vert \textit{\textbf{x}}) = \begin{cases} \frac{1}{2^{n-1}}&\bigoplus\limits_{i}a_i = \prod\limits_{i} x_i\\0&\text{otherwise.}\end{cases}
\end{equation}}
\end{definition}
\begin{definition}[$n$-Partite Even-Parity Box]
\label{def:ev}
\emph{An \emph{even-parity box for $n$ parties} is characterized by the following conditional distribution
\begin{equation}
P_n^{\text{c}}(\textit{\textbf{a}}\vert \textit{\textbf{x}}) = \begin{cases} \frac{1}{2^{n-1}}&\bigoplus\limits_{i}a_i = 0\\0&\text{otherwise.}\end{cases}
\end{equation}}
\end{definition}

Note that the box of Definition \ref{def:ev} is \emph{local}.
A convex combination of the boxes of Definitions \ref{def:pr} and \ref{def:ev} is called a \emph{correlated non-local box for n parties}.
\begin{definition}[Correlated Non-Local Boxes]
\emph{The \emph{family of correlated non-local boxes for n parties} is defined by
\begin{equation}
P^{\text{PR}}_{n,\varepsilon} = \varepsilon P^{\text{PR}}_n + (1-\varepsilon) P^{\text{c}}_n\ ,
\end{equation}
where $0\leq\varepsilon\leq1$.}
\end{definition}

\subsection{Communication as Systems}
In the protocols below, we will not only use $n$-partite boxes as resources, but also communication between some of the parties, \emph{i.e.}, signaling systems. This partial communication can be seen as a directed graph $G$ with $n$ vertices and directed edges which correspond to the one-way communication channel between the $n$ parties. We denote the one-way communication channels with $C(G)$, these channels can be used once in arbitrary order.

\section{A Reduction Calculus for Systems}
\subsection{Protocols}
A \emph{protocol} is a distributed algorithm that takes the inputs of the parties and produces outputs for every one. If the protocol also takes shared systems to produce outputs, it is called a \emph{reduction protocol}. Its goal can be to simulate some target system $T$, either perfectly or arbitrarily precisely \cite{forster}.
Assume there are $n$ parties that share~$m$ $n$-partite systems $S_1, S_2, ..., S_m$ and a random variable $R$. The parties get the input $\textit{\textbf{x}} = (x_1, x_2, ..., x_n)$, and finally, they output $\textit{\textbf{a}} = (a_1, a_2, ..., a_n)$. During the protocol, the parties are allowed to apply any classical circuitry to their local parts of the shared system. Such a circuitry is called \emph{wiring} and consists of choices for the inputs of the boxes and the generation of the outputs \cite{BLMPPR05, SPG06}.
\begin{definition}[Adaptive Protocol]
\emph{In an \emph{adaptive protocol}, every Party $i$ gets the input $x_i$ and acts as follows:
Party $i$ inputs $f_j(x_i, R, b_{i_1}, b_{i_2},...,b_{i_{j-1}})$ to the shared system $S_{i_j}$ for all $j\in \{1, 2, ..., m\}$, where the index $i_j$ depends on $x_i$, $R$, and the former output bits $b_{i_1}, b_{i_2},..., b_{i_{j-1}}$. The system $S_{i_j}$ outputs $b_{i_j}$ to party $i$. The final output of Party $i$ is given by the function $f^{x_i}(R, b_1, b_2, ..., b_m)$.
}
\end{definition}

\begin{definition}[Non-Adaptive Protocol]
\emph{In a \emph{non-adaptive protocol}, every Party $i$ gets the input $x_i$ and acts as follows:
Party $i$ inputs $f_j(x_i, R)$ to the shared system $S_j$ for all $j\in \{1, 2, ..., m\}$. The system $S_{j}$ outputs $b_{j}$ to party $i$. The final output of Party $i$ is given by the function $f^{x_i}(R, b_1, b_2, ..., b_m)$.}
\end{definition}

In contrast to adaptive protocols, no input of a system depends on the output of another one in a non-adaptive protocol.

\subsection{Resources Inequalities}
In the following, we use \emph{resources inequalities} as introduced in \cite{DHW03, DHW05, gpsww}. They are used to express whether some resource can be simulated by other resources plus shared randomness. Assume we have two systems $R$ and $R'$. We write
\begin{equation}
\label{reduction}
R \succeq R'
\end{equation}
if there exists a protocol that simulates $R'$ using $R$ and shared randomness. 

Clearly, if (\ref{reduction}) holds, then there also exists a protocol that simulates $R'$ using arbitrarily many copies of $R$ ($k$ copies of $R$ is written as $R^{\otimes k}$), an arbitrary other resource $R''$, and shared randomness
\begin{equation}
\lbrace R^{\otimes k}, R''\rbrace \succeq R'\ ,
\end{equation}
where $k\in \mathbb{N}\cup \{\infty \}$.

We write
\begin{equation}
R \succeq^* R'
\end{equation}
if there exists a protocol that simulates $R'$ using arbitrary many copies of $R$ and shared randomness. If $R'$ can be simulated arbitrarily precisely with a small number of copies of $R$ then we write
\begin{equation}
R \rightarrow^* R'\ .
\end{equation}

\subsection{Examples of Reductions}
With this notation, we are able to rephrase some well-known results. Obviously, 
\begin{equation}
\emptyset \succeq P
\end{equation}
if and only if $P$ is local.

From the definition of correlated non-local boxes for $n$ parties, we know that such a box is a convex combination of the even-parity box $P^{\text{c}}_n$ and the $n$-PR box $P^{\text{PR}}_n$. Since the even parity box is local, $\emptyset \succeq P^{\text{c}}_n$ and, therefore,
\begin{equation}
P_{n,\varepsilon'}^{\text{PR}} \succeq P^{\text{PR}}_{n, \varepsilon} \quad \text{for all $0\leq\varepsilon\leq\varepsilon' \leq 1$}\ .
\end{equation}
In a Section IV, we see that for every $0 < \varepsilon <1$ exists $\varepsilon' > \varepsilon$ such that 
\begin{equation}
{P_{n,\varepsilon}^{\text{PR}}}^{\otimes 2} \succeq P^{\text{PR}}_{n, \varepsilon'}\ ,
\end{equation}
and for all $0 < \varepsilon <1$
\begin{equation}
{P_{n,\varepsilon}^{\text{PR}}} \rightarrow^* P^{\text{PR}}_n\ .
\end{equation}

\section{Multi-Party Non-Locality Distillation}
Non-locality distillation protocols are executed by $n$ parties without communication. The protocol simulates a binary input/output system by classical (local) operations on non-local boxes \cite{FWW09}. The goal is to use weak non-local boxes for simulating stronger ones. Since these protocols only use a given set of boxes and local operations that can be simulated by shared randomness, we can describe the result of the non-locality distillation as a resources inequality: Assume that the distillation protocol uses as resources the boxes $P_1, P_2, ...,P_n$, where $n \in \mathbb{N}\cup \{\infty\}$, to simulate the box $P$. Therefore, we get the resources inequality
\begin{equation}
\lbrace P_1, P_2, ..., P_n\rbrace \succeq P\ .
\end{equation}

Brunner and Skrzypczyk \cite{BS09} proposed an adaptive protocol for two parties that distills non-locality in the asymptotic limit: All correlated non-local boxes are distilled arbitrarily closely to the (maximally non-local) PR box. In the notation of resources inequalities, we could describe this kind of distillation as
\begin{equation}
{P^{\text{PR}}_{2, \varepsilon}}^{\otimes 2} \succeq P^{\text{PR}}_{2, \varepsilon'} 
\end{equation}
and
\begin{equation}
{P^{\text{PR}}_{2, \varepsilon}} \rightarrow^* P^{\text{PR}}_2\ ,
\end{equation}
where $0<\varepsilon<1$ and $\varepsilon' = \varepsilon / 2\cdot (3 - \varepsilon )>\varepsilon$.
We extend this to all $n$-partite PR boxes in Protocol~\ref{prot} and Theorem \ref{thm:bsprot}.

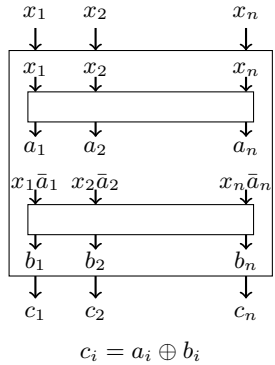
\begin{figure}[!htb]
\begin{center}
\begin{tikzpicture}
\begin{small}
\node [draw, rectangle, minimum width=3.5cm, minimum height=3cm] (A) at (0,0) {};
\node [draw, rectangle, minimum width=3cm, minimum height=0.4cm] (B)  at (0,0.75) {};
\node [draw, rectangle, minimum width=3cm, minimum height=0.4cm] (C) at (0,-0.75) {};
\draw[->, thick] (-1.4,1.8) to[] (-1.4,1.5);
\draw[->, thick] (-0.6,1.8) to[] (-0.6,1.5);
\draw[->, thick] (1.4,1.8) to[] (1.4,1.5);
\node [] () at (-1.4,2) {$x_1$};
\node [] () at (-0.6,2) {$x_2$};
\node [] () at (1.4,2) {$x_n$};

\draw[->, thick] (-1.4,1.15) to[] (-1.4,0.95);
\draw[->, thick] (-0.6,1.15) to[] (-0.6,0.95);
\draw[->, thick] (1.4,1.15) to[] (1.4,0.95);
\node [] () at (-1.4,1.25) {$x_1$};
\node [] () at (-0.6,1.25) {$x_2$};
\node [] () at (1.4,1.25) {$x_n$};

\draw[->, thick] (-1.4,0.55) to[] (-1.4,0.35);
\draw[->, thick] (-0.6,0.55) to[] (-0.6,0.35);
\draw[->, thick] (1.4,0.55) to[] (1.4,0.35);
\node [] () at (-1.4,0.2) {$a_1$};
\node [] () at (-0.6,0.2) {$a_2$};
\node [] () at (1.4,0.2) {$a_n$};

\draw[->, thick] (-1.4,-0.35) to[] (-1.4,-0.55);
\draw[->, thick] (-0.6,-0.35) to[] (-0.6,-0.55);
\draw[->, thick] (1.4,-0.35) to[] (1.4,-0.55);
\node [] () at (-1.4,-0.25) {$x_1\bar{a}_1$};
\node [] () at (-0.6,-0.25) {$x_2\bar{a}_2$};
\node [] () at (1.4,-0.25) {$x_n\bar{a}_n$};

\draw[->, thick] (-1.4,-0.95) to[] (-1.4,-1.15);
\draw[->, thick] (-0.6,-0.95) to[] (-0.6,-1.15);
\draw[->, thick] (1.4,-0.95) to[] (1.4,-1.15);
\node [] () at (-1.4,-1.3) {$b_1$};
\node [] () at (-0.6,-1.3) {$b_2$};
\node [] () at (1.4,-1.3) {$b_n$};

\draw[->, thick] (-1.4,-1.5) to[] (-1.4,-1.8);
\draw[->, thick] (-0.6,-1.5) to[] (-0.6,-1.8);
\draw[->, thick] (1.4,-1.5) to[] (1.4,-1.8);
\node [] () at (-1.4,-2) {$c_1$};
\node [] () at (-0.6,-2) {$c_2$};
\node [] () at (1.4,-2) {$c_n$};
\node [] () at (0,-2.5) {$c_i = a_i \oplus b_i$};

\end{small}
\end{tikzpicture}
\end{center}
\caption[$n$-PR Box destillation]{Generalized BS Protocol for $n$-PR boxes}
\label{fig:dbsprot}
\end{figure}

\begin{protocol}[Generalized BS Protocol for $n$-PR Boxes]
All n parties share two boxes, where we denote by $x_i$ the value that the $i$th party inputs to the first box and by $y_i$ the value that the $i$th party inputs to the second box. The output bit of the first box for the $i$th party is $a_i$, and the output bit of the second box is $b_i$. The n parties proceed as follows: $y_i =  x_i\bar{a}_i$ and they output, finally, $c_i = a_i \oplus b_i$ (see also Fig.~\ref{fig:dbsprot}).
\label{prot}
\end{protocol}

\begin{theorem}
\label{thm:bsprot}
Protocol \ref{prot} distills two copies of an arbitrary box $P^{\text{PR}}_{n,\varepsilon}$ with $0<\varepsilon<1$ to an $n$-partite correlated non-local box $P^{\text{PR}}_{n,\varepsilon'}$ with $\varepsilon'>\varepsilon$.
\begin{equation}
{P^{\text{\emph{PR}}}_{n, \varepsilon}}^{\otimes 2} \succeq P^{\text{\emph{PR}}}_{n, \varepsilon'}\ .
\end{equation} In the asymptotic limit of many copies, Protocol \ref{prot} distills any $P^{\text{PR}}_{n,\varepsilon}$ with $\varepsilon > 0$ to a box arbitrarily closely to the $n$-PR box
\begin{equation}
{P^{\text{\emph{PR}}}_{n, \varepsilon}} \rightarrow^* P^{\text{\emph{PR}}}_{n}\ .
\end{equation}
\end{theorem}

In the language of distillation, we say that in the asymptotic case of many copies, any $P^{\text{PR}}_{n,\varepsilon}$ with $\varepsilon > 0$ can be distilled arbitrarily closely to the $n$-PR box. This shows that also in the multipartite case, non-locality can be distilled.

\begin{IEEEproof}
We introduce the notation $A\triangleright B$, which means that the first box in Protocol \ref{prot} acts like $A$ and the second one like $B$.
The initial two-box state of Protocol \ref{prot} is given by
\begin{eqnarray}
P^{\text{PR}}_{n,\varepsilon} \triangleright P^{\text{PR}}_{n,\varepsilon}&=&\varepsilon^2 P^{\text{PR}}_{n}\triangleright P^{\text{PR}}_{n}\nonumber \\& & +~\varepsilon\left( 1-\varepsilon\right)\left( P^{\text{PR}}_{n} \triangleright P^{\text{c}}_{n}+ P^{\text{c}}_{n}\triangleright P^{\text{PR}}_{n} \right) \nonumber \\
& & +~(1-\varepsilon)^2 P^{\text{c}}_{n}\triangleright P^{\text{c}}_{n}\ .
\end{eqnarray}

We apply Protocol \ref{prot} and get the following relations: $P^{\text{PR}}_{n} \triangleright P^{\text{PR}}_{n} \equiv P^{\text{PR}}_{n}$ (\emph{i.e.}, $P^{\text{PR}}_n$ is a fixpoint), $P^{\text{PR}}_{n} \triangleright P^{\text{c}}_{n}\equiv P^{\text{PR}}_{n}$, $P^{\text{c}}_{n} \triangleright P^{\text{PR}}_{n}\equiv 2^{1-n}P^{\text{PR}}_{n} + \left( 1-2^{1-n}\right) P^{\text{c}}_{n}$, and $P^{\text{c}}_{n} \triangleright P^{\text{c}}_{n}\equiv P^{\text{c}}_{n}$.

After the application of Protocol \ref{prot}, we get the final box, which is
\begin{eqnarray}
P^{\text{PR}}_{n,\varepsilon'} & = & \frac{\varepsilon}{2^{n-1}}\left( 2^{n-1} + 1 - \varepsilon\right) P^{\text{PR}}_{n}\nonumber \\ & & + \left( 1 - \frac{\varepsilon}{2^{n-1}}\left( 2^{n-1} + 1 - \varepsilon\right)\right)  P^{\text{c}}_{n}\ .
\end{eqnarray}
Hence, $\varepsilon' = \varepsilon /2^{n-1}\cdot\left( 2^{n-1} + 1 - \varepsilon\right) $. We show that $\varepsilon' > \varepsilon$ for all $0< \varepsilon< 1$, therefore, the protocol takes any correlated non-local box $P^{\text{PR}}_{n,\varepsilon}$ to a stronger box $P^{\text{PR}}_{n,\varepsilon'}$.

We show that in the asymptotic regime of many copies, any $P^{\text{PR}}_{n,\varepsilon}$ with $0<\varepsilon<1$ can be distilled arbitrarily closely to the $n$-PR box. We are starting with $2^m$ copies of the box $P^{\text{PR}}_{n,\varepsilon}$ and get, finally, the box $P^{\text{PR}}_{n,\varepsilon_m}$, where
\begin{equation}T_n(\varepsilon) = \frac{\varepsilon}{2^{n-1}}\left( 2^{n-1} + 1 - \varepsilon\right)\ ,
\end{equation}
\begin{equation}
\varepsilon_m = T_n(\varepsilon_{m-1})\ , \quad \text{and} \quad \varepsilon_0 := \varepsilon\ .
\end{equation}
The fixed points of this map are $\varepsilon = 0$ and $\varepsilon = 1$. To analyze the stability of these two fixed points we calculate the eigenvalues of the Jacobian (since the map is one-dimensional, the Jacobian is a real value and not a matrix). For the box $P^{\text{c}}_{n}$ ($\varepsilon = 0$), we find $dT_n/d\varepsilon \vert_{\varepsilon=0} = 1 + 1/2^{n-1} > 1$, so this box is repulsive. For the other box $P^{\text{PR}}_{n}$ we find $dT_n/d\varepsilon\vert_{\varepsilon = 1} = 1 + 1/2^{n-1} - 1/2^{n-2} < 1$; the box is attractive.
\end{IEEEproof}

\section{Multi-Party Non-Locality Amplification}
The generalized BS protocol can be used to obtain non-locality amplification protocols for full-correlation boxes, where the use of communication is allowed to some of the parties. We allow a subset of the parties to use one-way communication channels (as often as required). We show that we are able to amplify a general class of full-correlation boxes arbitrarily closely to the maximum with such protocols.

\subsection{Construction of Full-Correlation Boxes}
\label{fullcor}
\begin{lemma}
\label{lem:boolf}
If $f$ is a Boolean function of the input elements $x_1,x_2,...,x_n$, then it can be written as
\begin{equation}
f(x_1,...,x_n) = \bigoplus\limits_{I \in \mathcal{I}} \left( a_I\cdot \bigwedge\limits_{i \in I}x_i \right)\ ,
\end{equation}
where $\mathcal{I} = \mathcal{P}\left( \lbrace 1,2,...,n \rbrace\right) $ and $a_I \in \lbrace 0, 1\rbrace$ for all $I \in \mathcal{I}$.
\end{lemma}

\begin{figure}[!htb]
\begin{center}
\begin{tikzpicture}
\begin{small}
\begin{scriptsize}
\node [draw, rectangle, minimum width=5cm, minimum height=1.5cm] (A) at (0,0) {};
\node [draw, rectangle, minimum width=1.3cm, minimum height=0.4cm] (B)  at (0,0) {3-PR Box};
\node [draw, rectangle, minimum width=1.3cm, minimum height=0.4cm] (C) at (1.575,0) {3-PR Box};
\node [draw, rectangle, minimum width=1.3cm, minimum height=0.4cm] (D) at (-1.575,0) {3-PR Box};
\end{scriptsize}
\node [draw, rectangle, minimum width=2cm, minimum height=1cm] (E) at (-5.5,0) {$1\oplus xy\oplus xz$};
\draw[->, thick] (-3,0) to[] (-4,0);
\draw[->, thick] (-4,0) to[] (-3,0);

\draw[->, thick] (-5.5,0.75) to[] (-5.5,0.5);
\draw[->, thick] (-6.1,0.75) to[] (-6.1,0.5);
\draw[->, thick] (-4.9,0.75) to[] (-4.9,0.5);
\node [] () at (-5.5,1) {$y$};
\node [] () at (-6.1,1) {$x$};
\node [] () at (-4.9,1) {$z$};

\draw[->, thick] (-5.5,-0.5) to[] (-5.5,-0.75);
\draw[->, thick] (-6.1,-0.5) to[] (-6.1,-0.75);
\draw[->, thick] (-4.9,-0.5) to[] (-4.9,-0.75);
\node [] () at (-5.5,-1) {$b$};
\node [] () at (-6.1,-1) {$a$};
\node [] () at (-4.9,-1) {$c$};

\draw[->, thick] (0,1) to[] (0,0.75);
\draw[->, thick] (1.575,1) to[] (1.575,0.75);
\draw[->, thick] (-1.575,1) to[] (-1.575,0.75);
\node [] () at (0,1.25) {$y$};
\node [] () at (1.575,1.25) {$z$};
\node [] () at (-1.575,1.25) {$x$};

\draw[->, thick] (0,-0.75) to[] (0,-1.3);
\draw[->, thick] (1.575,-0.75) to[] (1.575,-1);
\draw[->, thick] (-1.575,-0.75) to[] (-1.575,-1);
\node [] () at (0,-1.7) {$b = b_1\oplus b_2\oplus b_3$};
\node [] () at (1.575,-1.25) {$c = c_1\oplus c_2\oplus c_3$};
\node [] () at (-1.575,-1.25) {$a = a_1\oplus a_2\oplus a_3$};

\draw[->, thick] (0,0.4) to[] (0,0.2);
\draw[->, thick] (1.575,0.4) to[] (1.575,0.2);
\draw[->, thick] (-1.575,0.4) to[] (-1.575,0.2);
\node [] () at (0,0.55) {$y$};
\node [] () at (1.575,0.55) {$1$};
\node [] () at (-1.575,0.55) {$1$};

\draw[->, thick] (0,-0.2) to[] (0,-0.4);
\draw[->, thick] (1.575,-0.2) to[] (1.575,-0.4);
\draw[->, thick] (-1.575,-0.2) to[] (-1.575,-0.4);
\node [] () at (0,-0.55) {$b_2$};
\node [] () at (1.575,-0.55) {$b_3$};
\node [] () at (-1.575,-0.55) {$b_1$};

\draw[->, thick] (0.5,0.4) to[] (0.5,0.2);
\draw[->, thick] (2.075,0.4) to[] (2.075,0.2);
\draw[->, thick] (-1.075,0.4) to[] (-1.075,0.2);
\node [] () at (0.5,0.55) {$1$};
\node [] () at (2.075,0.55) {$z$};
\node [] () at (-1.075,0.55) {$1$};

\draw[->, thick] (0.5,-0.2) to[] (0.5,-0.4);
\draw[->, thick] (2.075,-0.2) to[] (2.075,-0.4);
\draw[->, thick] (-1.075,-0.2) to[] (-1.075,-0.4);
\node [] () at (0.5,-0.55) {$c_2$};
\node [] () at (2.075,-0.55) {$c_3$};
\node [] () at (-1.075,-0.55) {$c_1$};

\draw[->, thick] (-0.5,0.4) to[] (-0.5,0.2);
\draw[->, thick] (1.075,0.4) to[] (1.075,0.2);
\draw[->, thick] (-2.075,0.4) to[] (-2.075,0.2);
\node [] () at (-0.5,0.55) {$x$};
\node [] () at (1.075,0.55) {$x$};
\node [] () at (-2.075,0.55) {$1$};

\draw[->, thick] (-0.5,-0.2) to[] (-0.5,-0.4);
\draw[->, thick] (1.075,-0.2) to[] (1.075,-0.4);
\draw[->, thick] (-2.075,-0.2) to[] (-2.075,-0.4);
\node [] () at (-0.5,-0.55) {$a_2$};
\node [] () at (1.075,-0.55) {$a_3$};
\node [] () at (-2.075,-0.55) {$a_1$};

\end{small}
\end{tikzpicture}
\end{center}
\caption[Construction of the $1\oplus xy \oplus xz$-Box]{Construction of the $1\oplus xy \oplus xz$-Box}
\label{fig:equivalenz}
\end{figure}
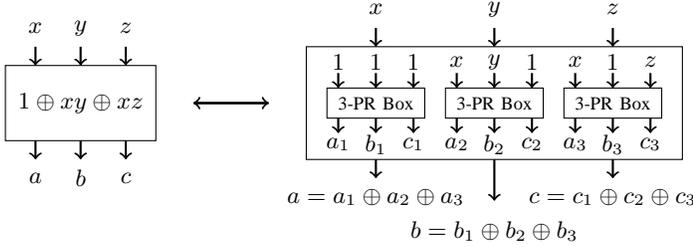

Hence, it is obvious that the full-correlation box associated to the Boolean function $f$ can be constructed by $\sum_{I \in \mathcal{I}}a_I$ $n$-PR boxes. Indeed, for every $a_I = 1$, an $n$-PR box is needed, where the $i$th party inputs $x_i$ if $i \in I$, and otherwise he inputs~$1$. Then, the box will output $b_i^I$. In the end, every party outputs $c_i = \bigoplus_{I \in \mathcal{I},\ a_I =1}b_i^I$. For an example, see Fig.~\ref{fig:equivalenz}. Note that the $n$-PR boxes belonging to $a_I$ where $\vert I \vert \leq 1$ are local and can be simulated by local operations and shared randomness.

We already know that all $n$-partite full-correlation boxes can be simulated by $n$-partite PR boxes. We define the set of all $n$-PR boxes that are needed to simulate the full-correlation box: Let

\begin{equation}\label{equ:j}
\mathcal{J} := \lbrace I \in \mathcal{I}\, \vert\, a_I = 1 \text{ and } \vert I\vert \geq 2\rbrace\ .
\end{equation}
This set can be partitioned into pairwise disjoint subsets $\lbrace J_1, J_2, ..., J_{n_\mathcal{J}}\rbrace$ such that all $A \in J_i$ and $B\in J_j$ fulfill $A\cap B = \emptyset$ for all $i \neq j$. We define the maximal number of such subsets as $n_{\mathcal{J}}$ and denote this partition as the \emph{empty-overlap partition of $\mathcal{J}$}.
We define, for all $I \in \mathcal{J}$, $m_I := \vert I \setminus \bigcup_{J \in \mathcal{J}\setminus I} J \vert$, \emph{i.e.}, the number of variables that only appear in the non-local box corresponding to $I \in \mathcal{J}$.

We take two  full-correlation boxes. The first is given by
\begin{equation}
P_1(a_1 \cdots a_{k_2}\vert x_1 \cdots x_{k_2}) = \begin{cases}\frac{1}{2^{k_2-1}}& \bigoplus\limits_{i=1}^{k_2} a_i = g_1(x_1,..., x_{k_2})\\0&\text{otherwise,}\end{cases}
\end{equation} 
where $g_1$ is a Boolean function which depends on all of its input variables, and $k_2<n$. The second box is defined as
\begin{equation}
P_2(b_{k_1}\cdots b_n\vert x_{k_1}\cdots x_n) = \begin{cases}\frac{1}{2^{n-k_1}}& \bigoplus\limits_{i=k_1}^{n} b_i = \prod\limits_{i=k_1}^{k_3}x_i \\0&\text{otherwise,}\end{cases}
\end{equation} 
where $0<k_1<k_2<k_3\leq n$. We construct an $n$-partite full-correlation box with these two boxes by taking the XOR of the two outputs $a_i$ and $b_i$ if Party $i$ participates at both boxes, otherwise the party outputs $a_i$ or $b_i$:
\begin{equation}
\label{constbox}
c_i = \begin{cases}a_i&  i \in\lbrace 1,2,..., k_1-1\rbrace \\ a_i \oplus b_i & i \in\lbrace k_1,k_1+1,..., k_2\rbrace \\b_i &  i \in\lbrace k_2+1, k_2+2,..., n\rbrace . \end{cases}
\end{equation}

\begin{lemma}
\label{lem:constr}
Box (\ref{constbox}) is equal (\emph{i.e.}, the joint probabilities are equal) to the full-correlation box defined by
\begin{equation}
P(\textbf{\textit{c}} \vert \textbf{\textit{x}}) = \begin{cases}\frac{1}{2^{n-1}}& \bigoplus\limits_{i=1}^n c_i = g_1(x_1,...,x_{k_2}) \oplus \prod\limits_{i=k_1}^{k_3}x_i \\0&\text{otherwise.}\end{cases}
\end{equation} 
\end{lemma}

\begin{IEEEproof}
The statement follows directly from the property of the full-correlation box that the set of outputs of any subset of $n-1$ parties (or smaller) is completely random \cite{BP05}, and the property that the XOR conserves randomness in case of independence.
\end{IEEEproof}

\begin{theorem}[Construction of a Full-Correlation Box]
\label{thm:full}
Let $P^f$ be the full-correlation box associated to the Boolean function $f$, and let $f$  be written as in Lemma \ref{lem:boolf}. If $f$ fulfills $n_\mathcal{J}=1$, then there exist subsets of parties such that the full-correlation box can be simulated with generalized PR boxes shared between the parties of a subset with the condition that the number of PR boxes in that some parties inputs all the time a constants is at most one.
\end{theorem}

\begin{IEEEproof}
We replace full-correlation boxes with $a_I = 1$ for $\vert I\vert \leq 1$ by the full-correlation box with $a_I = 0$ for $\vert I\vert \leq 1$, and all other $a_I$ for all $I \in \mathcal{I}\setminus \lbrace\emptyset\rbrace$ keep their values (\emph{i.e.}, we ignore the \emph{trivial part of the box}). We can do this by taking the XOR of the original box and the local box with $a_I = 1$ for $\vert I\vert \leq 1$. To get our original box back in the end, we take again the XOR of the modified box and the local box.

We replace the boxes step by step.
In the first step, we are beginning with a $n$-PR box  with the associated set $I$. To that end, we are looking for another $n$-PR box with associated set $J$ such that $I\cap J \neq \emptyset$ (this is possible because of the assumption made). Because of Lemma~\ref{lem:constr}, we are able to replace these two boxes by two smaller boxes: We substitute the first box by an $\vert I \setminus J\vert$-PR box with inputs $I$. The second box is substituted by an $(n- \vert I\vert )$-box, where we input $J$ and for the parties $\lbrace 1, 2, ..., n\rbrace \setminus (I\cup J)$, we input 1.

Assume that we have, in this way, replaced some $n$-PR boxes by new boxes. Let there be a further $n$-PR box which is not yet replaced, and whose input elements intersect with the input elements of the new box. We are making the same steps as before to replace these two boxes. In the end, we have replaced all $n$-PR boxes by a new box with the claimed properties. 
\end{IEEEproof}

\subsection{Imperfect Full-Correlation Boxes}
\label{ncfcb}
Assume we have a non-local full-correlation box $P^f$ associated to the Boolean function $f$ and  a local full-correlation box $P^{f_l}$ associated to the Boolean function
\begin{equation}
f_l = \bigoplus\limits_{\substack{I \in \mathcal{I}\setminus \mathcal{J} \\ a_I = 1}} \bigwedge\limits_{i \in I} x_i\  ,
\end{equation}
where $\mathcal{J}$ and the $a_i$'s are with respect to the function $f$.
This box corresponds to the trivial part of the full-correlation box $P^f$.

The imperfect box $P^f_\varepsilon$ is defined as the convex combination of these two boxes,
\begin{equation}
P^f_\varepsilon = \varepsilon P^f + (1-\varepsilon ) P^{f_l}\ ,
\end{equation}
where $0< \varepsilon < 1$.

We define the XOR of boxes:
\begin{definition}[XOR of boxes]
\emph{Let $P$ and $P'$ be two $n$-partite boxes that output $( a_1, a_2, ..., a_n)$, resp. $( b_1, b_2, ..., b_n)$, for the input $( x_1, x_2, ..., x_n)$.
The \emph{$XOR$ of the two boxes} $P$ and $P'$, \emph{i.e.}, $P \oplus P'$, is an $n$-partite box $P^*$ with output $( a_1\oplus b_1, a_2\oplus b_2, ..., a_n\oplus b_n)$ for the input $( x_1, x_2, ..., x_n)$.}
\end{definition}

\begin{definition}[$\text{XOR}^*$ of boxes]
\emph{Let $P_1$ and $P_2$ be two $n$-partite full correlation boxes, and $P_{i,\varepsilon}= \varepsilon P_1 + (1-\varepsilon)P^{\text{c}}$ for $i\in \{1, 2\}$. The \emph{$XOR^*$ of $P_{1,\varepsilon}$ and $P_{2,\varepsilon}$}, \emph{i.e.}, $P_{1,\varepsilon} \oplus^* P_{2,\varepsilon}$, is definded by
\begin{eqnarray}
P_{i, \varepsilon} \oplus^* P_{j, \varepsilon} & := & \varepsilon P_i\oplus P_j + (1-\varepsilon ) P^{\text{c}}\oplus P^{\text{c}}\nonumber\\ & = & \varepsilon P_i\oplus P_j + (1-\varepsilon ) P^{\text{c}}\ .
\end{eqnarray}}
\end{definition}

We can assume without loss of generality that $P^{f_l}$ is the even-parity box ($f_l = 0$, if this is not the case redefine $P^f_{\text{new}} = P^f \oplus P^{f_l}$, $P^{f_l}_{\text{new}} = P^{f_l} \oplus P^{f_l}$, and $P^f_{\varepsilon,\text{new}} = P^f _\varepsilon \oplus P^{f_l}$).
Note that the box $P^f$ can be written as the XOR of generalized $n$-PR boxes $P_1$, $P_2$,..., $P_m$ as seen in Section \ref{fullcor}
\begin{equation}
P^f = P_1 \oplus P_2 \oplus \cdots \oplus P_m\ .
\end{equation}
For that reason, $P^f_\varepsilon$ can be rewritten as
\begin{equation}
P^f_\varepsilon = P_{1,\varepsilon} \oplus^* P_{2, \varepsilon}\oplus^* \cdots \oplus^* P_{m, \varepsilon}\ ,
\end{equation}
where $P_{i, \varepsilon} = \varepsilon P_{i} + (1 - \varepsilon) P^{\text{c}}$ for all $i \in \{1, 2, ..., m\}$. That means we can simulate the box $P^f_\varepsilon$ with imperfect full-correlation boxes that all work correctly at the same time or all work incorrectly at the same time.

\begin{theorem}[Construction of an Imperfect  F.-C. Box]
\label{thm:imp}
Let $0<\varepsilon<1$, let $P^f$ be a full-correlation box associated to the Boolean function $f$, let $f$  be written as in Lemma \ref{lem:boolf}, and let $P^f_\varepsilon$ be defined as above. If $f$ fulfills $n_\mathcal{J}=1$, then there exists subsets of parties such that the box $P^f_\varepsilon$ can be simulated with imperfect generalized PR-boxes shared between the parties of a subset with the condition that the number of imperfect PR boxes in that some parties inputs all the time a constants is at most one. If all these imperfect generalized PR boxes work at the same time correctly and at the same time incorrectly then the simulation is equivalent to the box $P^f_\varepsilon$.
\end{theorem}

\begin{IEEEproof}
The proof is similar to the proof of Theorem \ref{thm:full}.
\end{IEEEproof}

\subsection{Protocols Based on Partial Communication}
Assume we have an $n$-partite full-correlation box $P^f$ that is to be simulated by one-way communication channels and shared randomness. The question is: How many one-way communication channels do we need for simulating an $n$-partite full-correlation box? Theorem \ref{thm:commcl} answers this question.

\begin{theorem}[Number of Communication Channels]
\label{thm:commcl}
Let $f$ be the Boolean function associated to an $n$-partite full-correlation box $P^f$, and let $f$ be defined as in Lemma~\ref{lem:boolf}. The number $N^{scratch}_{comm}$ of one-way communication channels to simulate the full-correlation box from scratch is
	\begin{equation}
	N_{comm}^{scratch} =  \left|\bigcup\limits_{I\in \mathcal{J}} I \right| - n_{\mathcal{J}}\ .
	\end{equation} 
\end{theorem}

\begin{IEEEproof}
We first prove the statement for $n_{\mathcal{J}} = 1$ by induction. We ignore the local part of the Boolean function $f$ (\emph{i.e.}, the terms of single variables) and start with the case when the function $f$ depends on two variables. The case $\vert\mathcal{J}\vert =2$ is equivalent to a PR box. From \cite{PBS11}, we know that it can be simulated by one one-way communication channel. Now, we assume that the claim is true for $\vert\mathcal{J}\vert \leq n$. Assume further that we have a function with  $\vert\mathcal{J}\vert =n+1$ that still fulfills the assumption. We substitute $1$ for $x_i$, where $x_i$ is the input which is an element of a minimal number of elements of $\mathcal{J}$. This new function also fulfills the assumption of the theorem. We also know that $\vert\mathcal{J}\vert =n$ and, therefore, we need $n-1$ communication channels to simulate the associated box. We combine all these $n$ function values into one variable. The original function can be written with two variables. Therefore, we are back at the case $\vert\mathcal{J}\vert =2$. Together, we need $n$ one-way communication channels for simulating a function with $\vert\mathcal{J}\vert =n+1$.

Assume now $n_{\mathcal{J}} > 1$. We write the original full-correlation box as a combination of $n_{\mathcal{J}}$ other non-local full-correlation boxes and at most one local full-correlation box (that can be simulated by shared randomness). Each of these boxes belongs to one of the sets of the empty-overlap partition $\{J_1, J_2, ..., J_{n_{\mathcal{J}}}\}$ of $\mathcal{J}$. The full-correlation box that belongs to $J_i$ is defined by the function 
\begin{equation}
f_i\left( x_1, x_2, ..., x_n\right)  = \bigoplus\limits_{J \in J_i} \bigwedge\limits_{j \in J} x_j\ .
\end{equation}
From the first part of the proof, we know that we need $\left|\bigcup_{J\in J_i} J \right| - 1$ communication channels to simulate this box from scratch. Thus, we need to simulate all the $n_{\mathcal{J}}$ $n$-partite non-local full-correlation boxes, for which we need
\begin{equation}
N_{comm}^{scratch} =  \left|\bigcup\limits_{I\in \mathcal{J}} I \right| - n_{\mathcal{J}}
\end{equation}
communication channels.
\end{IEEEproof}

From Theorem \ref{thm:commcl}, we know that all parties that belong to one of the sets of  the empty-overlap partition of $\mathcal{J}$, say $J_i$, have to communicate directly or indirectly to one of these parties. Corollary~\ref{cor:commcl} follows from this property.

\begin{corollary}
\label{cor:commcl}
Let $f$ be the Boolean function associated to an $n$-partite full-correlation box $P^f$, and let $f$ be defined as in Lemma \ref{lem:boolf}. Then
\begin{equation}
C(G) \succeq P^f\ ,
\end{equation}
where $G$ is a directed graph with $n$ vertices and the property that for every set $J_i$, \emph{i.e.}, a set of the empty-overlap partition of $\mathcal{J}$, there exists a vertex $v \in \left( \bigcup_{J\in J_i} J\right) $ such that from every other vertex $w \in \left( \bigcup_{J\in J_i} J\right) $, there exists a path to $v$ for all $i \in \{1,2,..., n_{\mathcal{J}}\}$.
\end{corollary}

\subsection{Protocol Based on Brunner/Skrypczyk-Protocol that Allows Partial Communication}
We have seen protocols that only use copies of some given boxes or partial communication. Now we study a combination of them.

Theorems \ref{thm:comm} and \ref{thm:comm2} state that a general class of full-correlation boxes can be simulated by (distillation) protocols and classical one-way communication channels. The number of these one-way channels is then smaller than the number of one-way communication channels we need if we do not apply a distillation protocol, \emph{i.e.}, operate from scratch. More specifically, there exists a minimal set of one-way communication channels that simulates such a full-correlation box, but only a subset of these channels is used to simulate the box using a (distillation) protocol.

Assume we have the non-local full-correlation box~$P^f$ associated to the Boolean function~$f$. Let the boxes~$P^{f_l}$  and~$P^{f}_\varepsilon$ be defined as in Section \ref{ncfcb}. We show that the box $P^f_\varepsilon$ can be distilled arbitrarily closely to the full-correlation box $P^f$ using partial communication if it fulfills certain conditions.

\begin{theorem}[Distillation with Communication I]
\label{thm:comm}
Let {$0<\varepsilon<1$}, let $P^f$ be a full-correlation box associated to the Boolean function $f$, let $f$  be written as in Lemma \ref{lem:boolf}, and let the box $P^f_\varepsilon$ be defined as in Section \ref{ncfcb} . If $f$ fulfills $n_\mathcal{J}=1$, then the number $N_{comm}^{distill}$ of one-way communication channels required for distilling  the box $P^f_\varepsilon$ up to the full-correlation box $P^f$ with using the generalized BS protocol is 
	\begin{equation}
	N_{comm}^{distill} \leq \begin{cases}n-1-\underset{I \in \mathcal{J}}{\max}(m_I)& \underset{I \in \mathcal{J}}{\max}(m_I) \neq n\\0& \underset{I \in \mathcal{J}}{\max}(m_I) = n. \end{cases}
	\end{equation} 
\end{theorem}

\begin{figure*}[t]
\begin{center}
\subfigure[]{\begin{tikzpicture}
\begin{small}
\begin{scriptsize}
\node [draw, rectangle, minimum width=7.2cm, minimum height=1.5cm] (A) at (0,0) {};
\node [draw, rectangle, minimum width=2cm, minimum height=0.4cm] (B)  at (0,0) {4-PR Box};
\node [draw, rectangle, minimum width=2cm, minimum height=0.4cm] (C) at (2.3,0) {4-PR Box};
\node [draw, rectangle, minimum width=2cm, minimum height=0.4cm] (D) at (-2.3,0) {4-PR Box};
\end{scriptsize}
\node [] () at (0,-1.55) {$a_i = a^1_i \oplus a^2_i \oplus a^3_i$};

\draw[red] (2.3,0) ellipse (1.2cm and 0.35 cm);
\node [red] () at (4,0) {local};

\draw[->, thick] (-2.7,1) to[] (-2.7,0.75);
\draw[->, thick] (-0.9,1) to[] (-0.9,0.75);
\draw[->, thick] (0.9,1) to[] (0.9,0.75);
\draw[->, thick] (2.7,1) to[] (2.7,0.75);
\node [] () at (-2.7,1.25) {$x_1$};
\node [] () at (-0.9,1.25) {$x_2$};
\node [] () at (0.9,1.25) {$x_3$};
\node [] () at (2.7,1.25) {$x_4$};

\draw[->, thick] (-2.7,-0.75) to[] (-2.7,-1);
\draw[->, thick] (-0.9,-0.75) to[] (-0.9,-1);
\draw[->, thick] (0.9,-0.75) to[] (0.9,-1);
\draw[->, thick] (2.7,-0.75) to[] (2.7,-1);
\node [] () at (-2.7,-1.25) {$a_1$};
\node [] () at (-0.9,-1.25) {$a_2$};
\node [] () at (0.9,-1.25) {$a_3$};
\node [] () at (2.7,-1.25) {$a_4$};

\draw[->, thick] (-0.75,0.4) to[] (-0.75,0.2);
\draw[->, thick] (-0.25,0.4) to[] (-0.25,0.2);
\draw[->, thick] (0.25,0.4) to[] (0.25,0.2);
\draw[->, thick] (0.75,0.4) to[] (0.75,0.2);
\node [] () at (-0.75,0.55) {$1$};
\node [] () at (-0.25,0.55) {$1$};
\node [] () at (0.25,0.55) {$x_3$};
\node [] () at (0.75,0.55) {$x_4$};

\draw[->, thick] (-0.75,-0.2) to[] (-0.75,-0.4);
\draw[->, thick] (-0.25,-0.2) to[] (-0.25,-0.4);
\draw[->, thick] (0.25,-0.2) to[] (0.25,-0.4);
\draw[->, thick] (0.75,-0.2) to[] (0.75,-0.4);
\node [] () at (-0.75,-0.55) {$a^2_1$};
\node [] () at (-0.25,-0.55) {$a^2_2$};
\node [] () at (0.25,-0.55) {$a^2_3$};
\node [] () at (0.75,-0.55) {$a^2_4$};

\draw[->, thick] (-3.05,0.4) to[] (-3.05,0.2);
\draw[->, thick] (-2.55,0.4) to[] (-2.55,0.2);
\draw[->, thick] (-2.05,0.4) to[] (-2.05,0.2);
\draw[->, thick] (-1.55,0.4) to[] (-1.55,0.2);
\node [] () at (-3.05,0.55) {$x_1$};
\node [] () at (-2.55,0.55) {$x_2$};
\node [] () at (-2.05,0.55) {$x_3$};
\node [] () at (-1.55,0.55) {$1$};

\draw[->, thick] (-3.05,-0.2) to[] (-3.05,-0.4);
\draw[->, thick] (-2.55,-0.2) to[] (-2.55,-0.4);
\draw[->, thick] (-2.05,-0.2) to[] (-2.05,-0.4);
\draw[->, thick] (-1.55,-0.2) to[] (-1.55,-0.4);
\node [] () at (-3.05,-0.55) {$a^1_1$};
\node [] () at (-2.55,-0.55) {$a^1_2$};
\node [] () at (-2.05,-0.55) {$a^1_3$};
\node [] () at (-1.55,-0.55) {$a^1_4$};

\draw[->, thick] (3.05,0.4) to[] (3.05,0.2);
\draw[->, thick] (2.55,0.4) to[] (2.55,0.2);
\draw[->, thick] (2.05,0.4) to[] (2.05,0.2);
\draw[->, thick] (1.55,0.4) to[] (1.55,0.2);
\node [] () at (3.05,0.55) {$1$};
\node [] () at (2.55,0.55) {$1$};
\node [] () at (2.05,0.55) {$1$};
\node [] () at (1.55,0.55) {$x_1$};

\draw[->, thick] (3.05,-0.2) to[] (3.05,-0.4);
\draw[->, thick] (2.55,-0.2) to[] (2.55,-0.4);
\draw[->, thick] (2.05,-0.2) to[] (2.05,-0.4);
\draw[->, thick] (1.55,-0.2) to[] (1.55,-0.4);
\node [] () at (3.05,-0.55) {$a^3_4$};
\node [] () at (2.55,-0.55) {$a^3_3$};
\node [] () at (2.05,-0.55) {$a^3_2$};
\node [] () at (1.55,-0.55) {$a^3_1$};

\end{small}
\end{tikzpicture}}
\subfigure[]{\begin{tikzpicture}
\begin{small}
\begin{scriptsize}
\node [draw, rectangle, minimum width=7.2cm, minimum height=1.5cm] (A) at (0,0) {};
\node [draw, rectangle, minimum width=1.5cm, minimum height=0.4cm] (B)  at (-0.2,0) {2-PR Box};
\node [draw, rectangle, minimum width=2cm, minimum height=0.4cm] (C) at (2.3,0) {4-PR Box};
\node [draw, rectangle, minimum width=1.5cm, minimum height=0.4cm, fill=yellow] (D) at (-2.1,0) {3-PR Box};
\end{scriptsize}
\node [] () at (0,-1.55) {$a_i = a^1_i \oplus a^2_i \oplus a^3_i$};

\draw[red] (2.3,0) ellipse (1.2cm and 0.35 cm);
\node [red] () at (4,0) {local};

\draw[->, thick] (-2.7,1) to[] (-2.7,0.75);
\draw[->, thick] (-0.9,1) to[] (-0.9,0.75);
\draw[->, thick] (0.9,1) to[] (0.9,0.75);
\draw[->, thick] (2.7,1) to[] (2.7,0.75);
\node [] () at (-2.7,1.25) {$x_1$};
\node [] () at (-0.9,1.25) {$x_2$};
\node [] () at (0.9,1.25) {$x_3$};
\node [] () at (2.7,1.25) {$x_4$};

\draw[->, thick] (-2.7,-0.75) to[] (-2.7,-1);
\draw[->, thick] (-0.9,-0.75) to[] (-0.9,-1);
\draw[->, thick] (0.9,-0.75) to[] (0.9,-1);
\draw[->, thick] (2.7,-0.75) to[] (2.7,-1);
\node [] () at (-2.7,-1.25) {$a_1$};
\node [] () at (-0.9,-1.25) {$a_2$};
\node [] () at (0.9,-1.25) {$a_3$};
\node [] () at (2.7,-1.25) {$a_4$};

\draw[->, thick] (0.175,0.4) to[] (0.175,0.2);
\draw[->, thick] (-0.9375,0.55) to[out=0,in=90] (-0.575,0.2);
\draw[->, thick] (2.05,0.4) to[] (2.05,0.2);
\node [] () at (0.175,0.55) {$x_4$};
\node [] () at (-1.0875,0.55) {$x_3$};

\draw[->, thick] (0.175,-0.2) to[] (0.175,-0.4);
\draw[->, thick] (-0.575,-0.2) to[] (-0.575,-0.4);
\draw[->, thick] (2.05,0.4) to[] (2.05,0.2);
\node [] () at (0.175,-0.55) {$a_4^1$};
\node [] () at (-0.575,-0.55) {$a_3^2$};

\draw[->, thick] (-2.6,0.4) to[] (-2.6,0.2);
\draw[->, thick] (-2.1,0.4) to[] (-2.1,0.2);
\draw[->, thick] (-1.23875,0.55) to[out=180,in=90] (-1.6,0.2);
\node [] () at (-2.6,0.55) {$x_1$};
\node [] () at (-2.1,0.55) {$x_2$};

\draw[->, thick] (-2.6,-0.2) to[] (-2.6,-0.4);
\draw[->, thick] (-2.1,-0.2) to[] (-2.1,-0.4);
\draw[->, thick] (-1.6,-0.2) to[] (-1.6,-0.4);
\node [] () at (-2.6,-0.55) {$a_1^1$};
\node [] () at (-2.1,-0.55) {$a_2^1$};
\node [] () at (-1.6,-0.55) {$a_3^1$};

\draw[->, thick] (3.05,0.4) to[] (3.05,0.2);
\draw[->, thick] (2.55,0.4) to[] (2.55,0.2);
\draw[->, thick] (2.05,0.4) to[] (2.05,0.2);
\draw[->, thick] (1.55,0.4) to[] (1.55,0.2);
\node [] () at (3.05,0.55) {$1$};
\node [] () at (2.55,0.55) {$1$};
\node [] () at (2.05,0.55) {$1$};
\node [] () at (1.55,0.55) {$x_1$};

\draw[->, thick] (3.05,-0.2) to[] (3.05,-0.4);
\draw[->, thick] (2.55,-0.2) to[] (2.55,-0.4);
\draw[->, thick] (2.05,-0.2) to[] (2.05,-0.4);
\draw[->, thick] (1.55,-0.2) to[] (1.55,-0.4);
\node [] () at (3.05,-0.55) {$a^3_4$};
\node [] () at (2.55,-0.55) {$a^3_3$};
\node [] () at (2.05,-0.55) {$a^3_2$};
\node [] () at (1.55,-0.55) {$a^3_1$};

\end{small}
\end{tikzpicture}}
\subfigure[]{\begin{tikzpicture}
\begin{small}
\begin{scriptsize}
\node [draw, rectangle, minimum width=5.5cm, minimum height=1.5cm] (A) at (0,0) {};
\node [draw, rectangle, minimum width=2cm, minimum height=0.4cm] (B)  at (-1.375,0) {original Box};
\node [draw, rectangle, minimum width=2cm, minimum height=0.4cm] (C) at (1.375,0) {4-PR Box};
\end{scriptsize}
\node [] () at (0,-1.7) {$a_i = a^1_i \oplus a^2_i$};

\draw[red] (1.375,0) ellipse (1.2cm and 0.35 cm);
\node [red] () at (3.15,0) {local};

\draw[->, thick] (-2.0625,1) to[] (-2.0625,0.75);
\draw[->, thick] (-0.6875,1) to[] (-0.6875,0.75);
\draw[->, thick] (0.6875,1) to[] (0.6875,0.75);
\draw[->, thick] (2.0625,1) to[] (2.0625,0.75);
\node [] () at (-2.0625,1.25) {$x_1$};
\node [] () at (-0.6875,1.25) {$x_2$};
\node [] () at (0.6875,1.25) {$x_3$};
\node [] () at (2.0625,1.25) {$x_4$};

\draw[->, thick] (-2.0625,-0.75) to[] (-2.0625,-1);
\draw[->, thick] (-0.6875,-0.75) to[] (-0.6875,-1);
\draw[->, thick] (0.6875,-0.75) to[] (0.6875,-1);
\draw[->, thick] (2.0625,-0.75) to[] (2.0625,-1);
\node [] () at (-2.0625,-1.25) {$a_1$};
\node [] () at (-0.6875,-1.25) {$a_2$};
\node [] () at (0.6875,-1.25) {$a_3$};
\node [] () at (2.0625,-1.25) {$a_4$};

\draw[->, thick, green] (1.9,-1.25) to[out=200,in=340] (0.825,-1.25);

\draw[->, thick] (1.625,0.4) to[] (1.625,0.2);
\draw[->, thick] (2.125,0.4) to[] (2.125,0.2);
\draw[->, thick] (1.125,0.4) to[] (1.125,0.2);
\draw[->, thick] (0.625,0.4) to[] (0.625,0.2);
\node [] () at (0.625,0.55) {$x_1$};
\node [] () at (1.125,0.55) {$1$};
\node [] () at (1.625,0.55) {$1$};
\node [] () at (2.125,0.55) {$1$};

\draw[->, thick] (1.625,-0.2) to[] (1.625,-0.4);
\draw[->, thick] (2.125,-0.2) to[] (2.125,-0.4);
\draw[->, thick] (1.125,-0.2) to[] (1.125,-0.4);
\draw[->, thick] (0.625,-0.2) to[] (0.625,-0.4);
\node [] () at (0.625,-0.55) {$a_1^2$};
\node [] () at (1.125,-0.55) {$a_2^2$};
\node [] () at (1.625,-0.55) {$a_3^2$};
\node [] () at (2.125,-0.55) {$a_4^2$};

\draw[->, thick] (-1.625,0.4) to[] (-1.625,0.2);
\draw[->, thick] (-2.125,0.4) to[] (-2.125,0.2);
\draw[->, thick] (-1.125,0.4) to[] (-1.125,0.2);
\draw[->, thick] (-0.625,0.4) to[] (-0.625,0.2);
\node [] () at (-0.625,0.55) {$0$};
\node [] () at (-1.125,0.55) {$x_3$};
\node [] () at (-1.625,0.55) {$x_2$};
\node [] () at (-2.125,0.55) {$x_1$};

\draw[->, thick] (-1.625,-0.2) to[] (-1.625,-0.4);
\draw[->, thick] (-2.125,-0.2) to[] (-2.125,-0.4);
\draw[->, thick] (-1.125,-0.2) to[] (-1.125,-0.4);
\draw[->, thick] (-0.625,-0.2) to[] (-0.625,-0.4);
\node [] () at (-0.625,-0.55) {$a_4^1$};
\node [] () at (-1.125,-0.55) {$a_3^1$};
\node [] () at (-1.625,-0.55) {$a_2^1$};
\node [] () at (-2.125,-0.55) {$a_1^1$};

\end{small}
\end{tikzpicture}}
\end{center}
\caption{(a) Simulating the full-correlation box with three 4-PR boxes. (b) Simulation of the full-correlation box with generalized PR boxes without a constant input and a local box. (c) How to simulate the 3-PR box with the original full-correlation box and a local box.}
\label{fig:example}
\end{figure*}
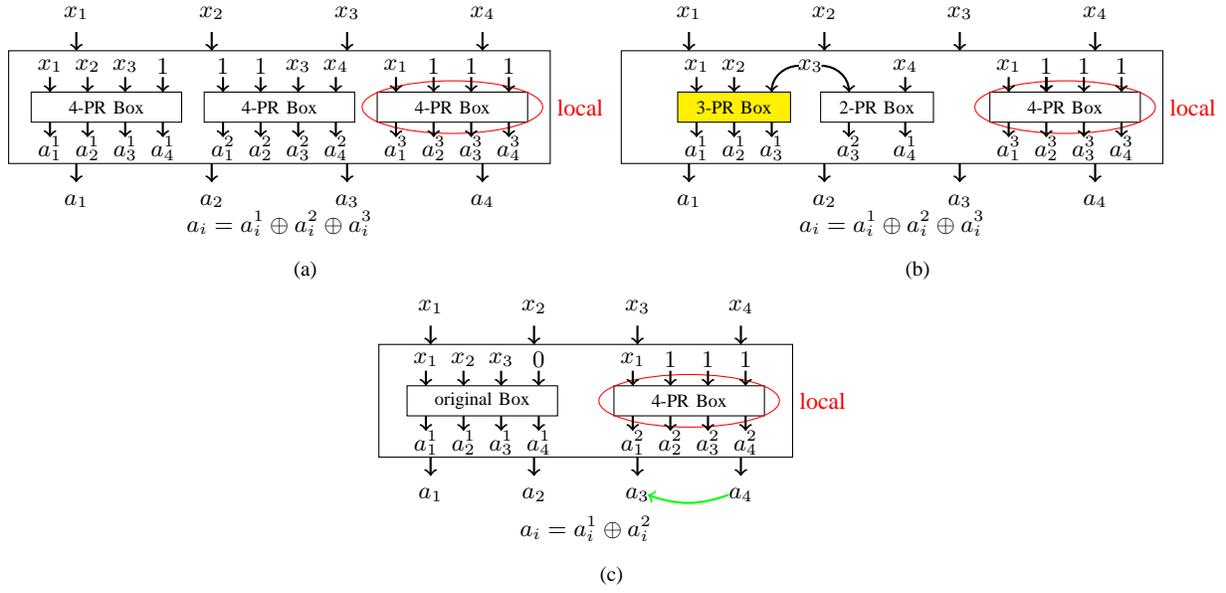

\begin{IEEEproof}
Here, we replace full-correlation boxes with $a_I = 1$ for $\vert I\vert \leq 1$ by the full-correlation box with $a_I = 0$ for $\vert I\vert \leq 1$, and the other $a_I$, for all $I \in \mathcal{I}\setminus \lbrace\emptyset\rbrace$, keep their values. We do the same with the imperfect full-correlation box $P^f_\varepsilon$. We can do this by taking the XOR of the original box and the local box with $a_I = 1$ for $\vert I\vert \leq 1$. To get our original box back in the end, we take again the XOR of the modified box and the local box.

We assume that the replacement is made according to Theorem \ref{thm:imp}. We have replaced the original correlated $n$-partite boxes in such a way that the correlated box with constant input does not correspond to the original correlated $n$-partite box belonging to the largest $m_I$. This is possible since we can replace this box first. We are now able to isolate the box belonging to the largest $m_I$. Therefore, we allow all parties that appear at least twice as well as the parties that input all the time a constant to communicate their inputs and outputs to a party that also has an input for the isolated box. We have isolated the correlated multipartite box belonging to the largest $m_I$, and we are able to apply the generalized BS protocol to this box. All the other correlated boxes that appear in the abstraction of Theorem \ref{thm:imp} can be simulated by the communication of the parties and shared randomness. So we will need $\max_{I \in \mathcal{J}}(m_I)$ one-way-communication channels less than if we started from scratch.
\end{IEEEproof}

The following is a corollary of Theorem \ref{thm:comm}:

\begin{corollary}
Let $0<\varepsilon<1$, let $P^f$ be a full-correlation box associated to the Boolean function $f$, let $f$  be written as in Lemma \ref{lem:boolf}, and let $I$ be the set of the inputs of the box that belongs to the largest $m_I$. If $f$ fulfills $n_\mathcal{J}=1$, then
\begin{equation}
\lbrace {P^f_\varepsilon}, C(G)\rbrace \rightarrow^* P^f\ ,
\end{equation}
where the box $P^f_\varepsilon$ is defined as in Section \ref{ncfcb} and $G$ is a directed graph with $n$ vertices with the property that there exists a vertex $v \in \left( \bigcup_{J\in \mathcal{J}} J\right) \cap I$ such that from every vertex $w \in \left( \{1, 2, ...,n\}\setminus I\right)  \cup \left( \bigcup_{J\in \mathcal{J}} J\right)$, there exists a path to $v$.
\end{corollary}

\begin{corollary}
Let $0<\varepsilon<1$, let $P^f$ be a full-correlation box associated to the Booelan function $f$, and let $f$  be written as in Lemma~\ref{lem:boolf}. If $n_\mathcal{J}=1$ and $\max_{I \in \mathcal{J}}(m_I) > n - \vert\bigcup_{I\in \mathcal{J}} I\, \vert$, then
\begin{equation}
N_{comm}^{distill} < N_{comm}^{scratch}\ ,
\end{equation}
where $N_{comm}^{distill}$ is the number of one-way communication channels needed for distilling the box $P^f_\varepsilon$ that is defined as in Section \ref{ncfcb}.
\label{cor:ungl}
\end{corollary}

\begin{IEEEproof}
The statement follows from Theorems~\ref{thm:commcl} and~\ref{thm:comm}.
\end{IEEEproof}

\begin{theorem}[Distillation with Communication II]
\label{thm:comm2}
Let $0<\varepsilon<1$, let $P^f$ be a full-correlation box associated to the Boolean function $f$, let $f$  be written as in Lemma \ref{lem:boolf}, and let the box $P^f_\varepsilon$ be defined as in Section \ref{ncfcb}. If 
\begin{equation}
\max_{I \in \mathcal{J}}(m_I)>n - \left|\bigcup_{I\in \mathcal{J}} I \right|\ ,
\end{equation} 
and $n_\mathcal{J}=1$, then there exists a graph $G$ with $N_{comm}^{scratch}$ directed edges and a proper subgraph $G' \subset G$ with $N_{comm}^{distill}$ directed edges such that $C(G)\succeq P^f$ and $\lbrace {P^f_\varepsilon}, C(G')\rbrace \rightarrow^* P^f$.
\end{theorem}

\begin{IEEEproof}
The statement follows from Theorems~\ref{thm:commcl} and~\ref{thm:comm}, and Corollary \ref{cor:commcl}.
\end{IEEEproof}

All extremal three-partite full-correlation boxes of the non-signalling polytope fulfill the conditions of Corollary~\ref{thm:comm2}. For more parties, it is unknown how many extremal boxes also fulfill the condition.

\section{Examples}

\subsection{Example of an Amplifiable System}
In this example, we simulate the following full-correlation box: 
\begin{equation}
P^1(\textit{\textbf{a}}\vert \textit{\textbf{x}}) = \begin{cases} \frac{1}{2^3}&\text{$\bigoplus\limits_{i=1}^{4} a_i = x_1x_2x_3\oplus x_3x_4\oplus x_1$}\\0&\text{otherwise.}\end{cases}
\end{equation}

Therefore, we determine first the above-defined sets and constants. Let $\mathcal{I}= \mathcal{P}(\lbrace 1, 2, 3, 4\rbrace )$. From Lemma~\ref{lem:boolf}, we know that all $a_I = 1$ for $I\in \lbrace\lbrace 1,2,3\rbrace , \lbrace 3,4\rbrace , \lbrace 1\rbrace\rbrace$, and otherwise $a_I = 0$. This means that the given full-correlation box can be simulated by three 4-PR boxes with some constant inputs, where one of these boxes is local (see Fig. \ref{fig:example} a)).
We are also able to determine the set $\mathcal{J}$ of non-local $n$-PR boxes that are required to simulate the full-correlation box:
\begin{equation}
\mathcal{J}= \lbrace\lbrace 1,2,3\rbrace , \lbrace 3,4\rbrace\rbrace
\end{equation}
Both of these non-local 4-PR boxes can be obtained from the original box by taking the XOR of the original box and the local 4-PR box when every party inputs his bits except for the parties that input the constant 1 to the 4-PR box, they input~0 in both boxes. If we apply Theorem~\ref{thm:comm} (i), then we know that the non-local part of the original full-correlation box can be simulated by two connected $n$-PR boxes with no constant input (see Fig. \ref{fig:example} b)).

Since there is only one set in the empty-overlap partition of $\mathcal{J}$, $n_{\mathcal{J}} = 1$. Therefore, the number of required one-way communication channels for simulating the full-correlation box can be calculated according to Theorem~\ref{thm:commcl}:
\begin{equation}
N_{comm}^{scratch} =  \left|\bigcup\limits_{I\in \mathcal{J}} I \right| - 1 = 3\ .
\end{equation}

One of the graphs that charaterizes the one-way communication channels is $G=(V,E)$ with $V=\{1, 2, 3, 4, 5\}$ and $E = \{(4,3), (3,2), (2,1)\}$. That leads to
\begin{equation}
C(G) \succeq P^1\ .
\end{equation}

Obviously, this box is not local. We define the trivial part of this full-correlation box

\begin{equation}
P^L(\textit{\textbf{a}}\vert \textit{\textbf{x}}) = \begin{cases} \frac{1}{2^{3}}&\text{$\bigoplus\limits_{i=1}^{4} a_i = x_1$}\\0&\text{otherwise.}\end{cases}
\end{equation}

We start with the second part of the example, where we show in detail how we take a box from the family $P_\varepsilon = \varepsilon P + (1-\varepsilon)P^L$, where $0<\varepsilon < 1$, to the box $P(\textit{\textbf{a}}\vert \textit{\textbf{x}})$. For that, we determine first which of the parties have to communicate. Therefore, we calculate the number of parties that only belong to one of the non-local 4-PR boxes: $m_{\lbrace 1,2,3\rbrace} = 2$ and
$m_{\lbrace 3,4\rbrace} =1$. This means that we isolate the box that belongs to the 4-PR box with three arbitrary inputs. This can be done in the same way as before: We input $(x_1, x_2, x_3, 0)$ to $P_\varepsilon$ and the local box and take the XOR of its outputs. Then, we use a one-way communication channel from Party~4 to~3. This corresponds to a graph $G' = (V',E')$ with $V'=V$ and $E'=\{(4,3)\}$, which means we need one one-way communication channel. Remember that the communication channel can be used as often as required. Hence, we are able to simulate perfectly the other 2-PR boxes, and the imperfect 3-PR box can be isolated by communicating the inputs and outputs of the 2-PR box to Party 3 (see Fig. \ref{fig:example} c)). We have isolated the box $P_{3,\varepsilon}^{PR}$ that is known to be asymptotically distillable up to $P_3^{PR}$ by the generalized BS protocol. In this way, we are able to take the box $P_\varepsilon$ to the full-correlation box in the beginning. This results in the resources inequality
\begin{equation}
{P^L}^{\otimes\infty} \otimes C(G') \succeq P^1\ .
\end{equation}

We get that $G'$ is a proper subgraph of $G$ and the number of one-way communication channels that is needed for this kind of protocol is $N^{distill}_{comm} = 1$, i.e., less than $N^{scratch}_{comm} = 3$.

\subsection{Example of a Non-Amplifiable System}
In this example we simulate the following full-correlation box: 
\begin{equation}
P^2(\textit{\textbf{a}}\vert \textit{\textbf{x}}) = \begin{cases} \frac{1}{2^{5}}&\text{$\bigoplus\limits_{i=1}^{6} a_i = f(x_1, x_2, ..., x_6) $}\\0&\text{otherwise,}\end{cases}
\end{equation}
where $ f(x_1, x_2, ..., x_6) = x_1x_2\oplus x_2x_3\oplus x_4x_5x_6\oplus x_5$.

Let $\mathcal{I}= \mathcal{P}(\lbrace 1, 2, 3, 4, 5, 6\rbrace )$. From Lemma~\ref{lem:boolf} we know that all $a_I = 1$ for $I\in \lbrace\lbrace 1,2\rbrace , \lbrace 2,3\rbrace , \lbrace 4,5,6\rbrace , \lbrace 5\rbrace\rbrace$, and otherwise $a_I = 0$. This means that the given full-correlation box can be simulated by four 6-PR boxes with some constant inputs, where one of these boxes is local.
We are also able to assign the set $\mathcal{J}$ of non-local $n$-PR boxes that are needed to simulate the full-correlation box:
\begin{equation}
\mathcal{J}= \lbrace\lbrace 1,2\rbrace , \lbrace 2,3\rbrace , \lbrace 4,5,6\rbrace\rbrace\ .
\end{equation}
Each of these three non-local 6-PR boxes can be obtained from the original box by taking the XOR of the original box and the local 5-PR box when every party inputs its bits except for the parties that input the constant 1 to the 5-PR box, they input 0 in both boxes.

Since we know $\mathcal{J}$, we can determine the empty-overlap partition $\{J_1, J_2\}$, where $J_1 = \{\{ 1, 2\}, \{ 2, 3\}\}$ and $J_2 = \{\{ 4, 5, 6\}\}$. Therefore, $n_{\mathcal{J}} = 2$ and the number of required one-way communication channels for simulating the full-correlation box can be calculated according to Theorem~\ref{thm:commcl}:
\begin{equation}
N_{comm}^{scratch} =  \left|\bigcup\limits_{I\in \mathcal{J}} I \right| - n_{\mathcal{J}} = 4\ .
\end{equation}

One of the graphs that charaterizes the one-way communication channels is $G=(V,E)$ with $V=\{1, 2, 3, 4, 5, 6\}$ and $E = \{(1,2), (2,3), (4,5), (5,6)\}$. That leads us to
\begin{equation}
C(G) \succeq P^2\ .
\end{equation}
Since $n_{\mathcal{J}} \neq 1$, Theorem~\ref{thm:comm} does not apply.

\section{Conclusion}
We have studied the problem of non-locality distillation in the
multi-partite setting. We have found, first, that
arbitrarily weakly non-local non-isotropic approximations to
the natural
generalization of a PR box to $n$ parties are distillable by an
adaptation of a protocol for two parties. Second, this can be
applied to showing that a much more general class of extremal
correlations, including \emph{all} purely three-partite correlations, can be amplified to using {\em partial\/} communication
requring only a subset of directed pairwise channels than as compared to the case when weak systems can be used. In this context, weak
non-locality, hence, manages to replace communication between a
subset of parties. It remains a challenging open problem to
understand,
classify, and apply multi-party non-locality systematically. It seems that for
certain tasks (such as randomness amplification \cite{galle, galli}), multi-party
non-locality
outperforms bipartite correlations.


%

%

\section*{Acknowledgment}
The authors thank {\"A}.~Baumeler, D.\ Frauchiger, A.\ Montina,  M.\ Pfaffhauser, J.\ Rashid, and B.\ Salwey for helpful discussions.

\end{document}